\documentclass[twocolumn,conference]{IEEEtran}
\usepackage[T1]{fontenc}
\usepackage[latin9]{inputenc}
\usepackage{color}
\usepackage{array}
\usepackage{verbatim}
\usepackage{float}
\usepackage{amsmath}
\usepackage{amsthm}
\usepackage{graphicx}
\usepackage[unicode=true,
 bookmarks=false,
 breaklinks=false,pdfborder={0 0 0},pdfborderstyle={},backref=false,colorlinks=true]
 {hyperref}
\hypersetup{pdftitle={Generating End-to-End Adversarial Examples for Malware Classifiers Using Explainability},
 pdfauthor={Ishai Rosenberg},
 pdfpagelayout=OneColumn, pdfnewwindow=true, pdfstartview=XYZ, plainpages=false}
\usepackage{breakurl}

\makeatletter

\providecommand{\tabularnewline}{\\}
\floatstyle{ruled}
\newfloat{algorithm}{tbp}{loa}
\providecommand{\algorithmname}{Algorithm}
\floatname{algorithm}{\protect\algorithmname}

\theoremstyle{plain}
\newtheorem{thm}{\protect\theoremname}
\theoremstyle{definition}
\newtheorem{defn}[thm]{\protect\definitionname}

\usepackage[caption=false,font=footnotesize]{subfig}

\usepackage{fancyhdr}
\makeatother

\providecommand{\definitionname}{Definition}
\providecommand{\theoremname}{Theorem}

\begin{document}

\title{Generating End-to-End Adversarial Examples for Malware Classifiers
Using Explainability }

\author{\IEEEauthorblockN{Ishai~Rosenberg and Shai~Meir and Jonathan~Berrebi and Ilay~Gordon
and Guillaume~Sicard and Eli~(Omid)~David}\IEEEauthorblockA{Deep Instinct Ltd\\
\{ishair, shaim, jonathanb, ilayc, guillaumes, david\}@deepinstinct.com}}
\maketitle
\begin{abstract}
In recent years, the topic of explainable machine learning (ML) has
been extensively researched. Up until now, this research focused on
regular ML users use-cases such as debugging a ML model. This paper
takes a different posture and show that adversaries can leverage explainable
ML to bypass multi-feature types malware classifiers. Previous adversarial
attacks against such classifiers only add new features and not modify
existing ones to avoid harming the modified malware executable's functionality.
Current attacks use a single algorithm that both selects which features
to modify and modifies them blindly, treating all features the same.
In this paper, we present a different approach. We split the adversarial
example generation task into two parts: First we find the importance
of all features for a specific sample using explainability algorithms,
and then we conduct a feature-specific modification, feature-by-feature.
In order to apply our attack in black-box scenarios, we introduce
the concept of transferability of explainability, that is, applying
explainability algorithms to different classifiers using different
features subsets and trained on different datasets still result in
a similar subset of important features. We conclude that explainability
algorithms can be leveraged by adversaries and thus the advocates
of training more interpretable classifiers should consider the trade-off
of higher vulnerability of those classifiers to adversarial attacks.
\end{abstract}

\IEEEpeerreviewmaketitle{}

\section{Introduction}

In recent years, the topic of explainable and interpretable machine
learning (ML) has been extensively researched. Explainable machine
learning can be used by the ML model users and developers, e.g., to
debug prediction errors of an existing ML model on specific samples
or to provide human explanations of models' decisions by highlighting
the features that had the highest impact on a specific sample decision.

In this paper, we demonstrate the ability of an adversary to use the
explainability of multi-feature types malware classifiers algorithms
to produce the most important features for a known, white-box model.
Then, due to the concept of transferable explanations, the same important
features are relevant to the target black-box classifier. Therefore,
by modifying existing malware's important features by the white-box
model's explanation to fool it, not only the white-box model would
be fooled, but also the target black-box model.

The principle of transferability of adversarial examples, that is,
adversarial examples crafted against one model are also likely to
be effective against other models, is already known \cite{DBLP:journals/corr/SzegedyZSBEGF13,Goodfellow14,DBLP:journals/corr/PapernotMG16}
and was also evaluated in the cyber security domain \cite{DBLP:conf/raid/RosenbergSRE18}.
However, most research related to adversarial examples is focused
on adversarial examples containing a single feature type: changing
pixel's color in an input image, modifying words in an input sentence,
etc. In those cases, modifying each feature has the same level of
difficulty, because they all have the same feature type.

In contrast, in the cyber security domain there is a unique challenge
which is not addressed by previous research: Malware classifiers (which
gets an executable file as input and predict the labels of benign
or malicious for each file) often use more than a single feature type
(see Section \ref{subsec:There-Are-Many}). Thus, adversaries who
want to subvert those systems should consider modifying more than
a single feature type. Some feature types are easier to modify without
harming the executable functionality than others (see Section \ref{subsec:There-Are-Many}).
In addition, even the same feature type might be modified differently
depending on the sample format. For instance, modifying a printable
string inside a PE file might be more challenging than modifying a
word within the content of an email, although the feature type is
the same. This means that we should not only take into account the
impact of a feature on the prediction, but also the difficulty of
modifying this feature type \cite{KATZIR2018419}. In addition, some
features are dependent on other features, meaning that modifying one
feature requires modifying other features for the executable to continue
functioning. For instance, %
{} adding strings to the file (as done in \cite{DBLP:conf/raid/RosenbergSRE18})
will necessarily impact the features dependent on printable characters,
e.g., entropy.

The end result is that when adversaries want to modify a PE file without
harming its functionality, the feature modification must be done manually.
In this way, only features that are easy to modify, not dependent
on other features who are challenging to modify (which is feature
specific) would be modified. Thus, we would want to modify the smallest
numbers of features, because each feature's modification requires
a manual effort. Moreover, each modified feature can create a feature
distribution anomaly that could be detected by anomaly detection algorithms
(e.g., \cite{DBLP:journals/scn/ZakeriDA15}). Therefore, the adversary
aims to modify as little features as possible, even if he/she could
modify all features automatically. In order to achieve this goal,
the adversary would like to get the list of most impactful features
for a specific sample (the malware which tries to bypass the malware
classifier) and manually select the features that are the easiest
to modify. This is the approach we take in this paper, as opposed
to previous adversarial attacks, that try to do both in the same algorithm
and thus try to modify all features in the same manner, resulting
in generating malware executables that don't run. 

In order to select the most indicative features for a sample, the
adversary can use explainability algorithms. However, the adversary
is not familiar with the architecture of the target malware classifier.
In order to resolve this issue, he/she can train his/her own malware
classifier and use its features instead. Those features are likely
to have a high impact even in the target classifier, due to the concept
of transferability of explainability.

\subsection{\label{subsec:The-Challenges-in}The Challenges in End-To-End Adversarial
Examples of Malware Executables}

Most published adversarial attacks, including those that were published
at academic cyber security conferences have focused on the computer
vision domain, e.g., generating a cat image that would be classified
as a dog by the classifier. However, the cyber security domain \textendash{}
and particularly the malware detection task - seems a much more relevant
domain for adversarial attacks, because while in the computer vision
domain, there is no concrete adversary who wants cats to be classified
as dogs, in the cyber security domain, there are actual adversaries
with clear targeted goals. Examples include ransomware developers
who depend on the ability of their ransomware to evade anti-malware
products that would prevent both its execution and the developers
from collecting the ransom money, and other types of malware that
need to steal user information (e.g., keyloggers), spread across the
network (worms) or perform any other malicious functionality while
remaining undetected. 

Given the obvious relevance of the cyber security domain to adversarial
attacks, why do most adversarial learning researchers focus on computer
vision? Besides the fact that image recognition is a popular machine
learning research topic, another major reason is that performing an
end-to-end adversarial attack in the cyber security domain is more
difficult than performing such an attack in the computer vision domain.
The differences between adversarial attacks performed in those two
domains and the challenges that arise in the cyber security domain
are discussed in the subsections that follow.

\subsubsection{\label{subsec:Executable-(Malicious)-Functiona}Executable (Malicious)
Functionality}

Any adversarial executable must preserve its malicious functionality
after the sample\textquoteright s modification. This might be the
main difference between the image classification and malware detection
domains, and pose the greatest challenge. In the image recognition
domain, the adversary can change every pixel\textquoteright s color
(to a different valid color) without creating an \textquotedblleft invalid
picture\textquotedblright{} as part of the attack. However, in the
cyber security domain, modifying an API call or arbitrary executable\textquoteright s
content byte value might cause the modified executable to perform
a different functionality (e.g., modifying a WriteFile() call to ReadFile()
) or even crash (if you change an arbitrary byte in an opcode to an
invalid opcode that would cause an exception). 

In order to address this challenge, adversaries in the cyber security
domain must implement their own methods (which are usually feature-specific)
to modify features in a way that will not break the functionality
of the executable. For instance, the adversarial attack used in Rosenberg
et al. \cite{DBLP:conf/raid/RosenbergSRE18} modifies API call traces
in a functionality preserving manner.

\subsubsection{\label{subsec:There-Are-Many}There are Many Feature Types }

In the cyber security domain, classifiers usually use more than a
single feature type as input (e.g., malware detection using both PE
header metadata and byte entropy in Saxe et al. \cite{Saxe2015}).
Some feature types are easier to modify without harming the executable
functionality than others. For instance, in the adversarial attack
used in Rosenberg et al. \cite{DBLP:conf/raid/RosenbergSRE18}, appending
printable strings to the end of file is much easier than adding API
calls using a dedicated framework built for this purpose. In contrast,
in an image adversarial attack, modifying each pixel has the same
level of difficulty.

\subsubsection{Executables are More Complex than Images}

An image used as input to an image classifier (usually a convolutional
neural network, CNN) is represented as a fixed size matrix of pixel
colors. If the actual image has different dimensions than the input
matrix, the picture will usually be resized, clipped, or padded to
fit the dimension limits. 

An executable, on the other hand, has a variable length: executables
can range in size from several KB to several GB. It\textquoteright s
also unreasonable to expect a clipped executable to keep its original
classification. Let's assume we have a 100MB benign executable into
which we inject a shellcode at a function near the end-of-file. If
the shellcode is clipped in order to fit the malware classifier's
dimensions, there is no reason that the file would be classified as
malicious, because its benign variant would be clipped to the exact
same form.

In addition, the code execution path of an executable may depend on
the input, and thus, the adversarial perturbation should support any
possible input that the malware may encounter when executed in the
target machine.

While this is a challenge for malware classifier implementation, it
also affects adversarial attacks against malware classifiers. Attacks
in which you have a fixed input dimension, (e.g., a 28{*}28 matrix
for MNIST images), are much easier to implement than attacks in which
you need to consider the variable file size.

\begin{figure}[t]
\begin{centering}
\textsf{\includegraphics[scale=0.2]{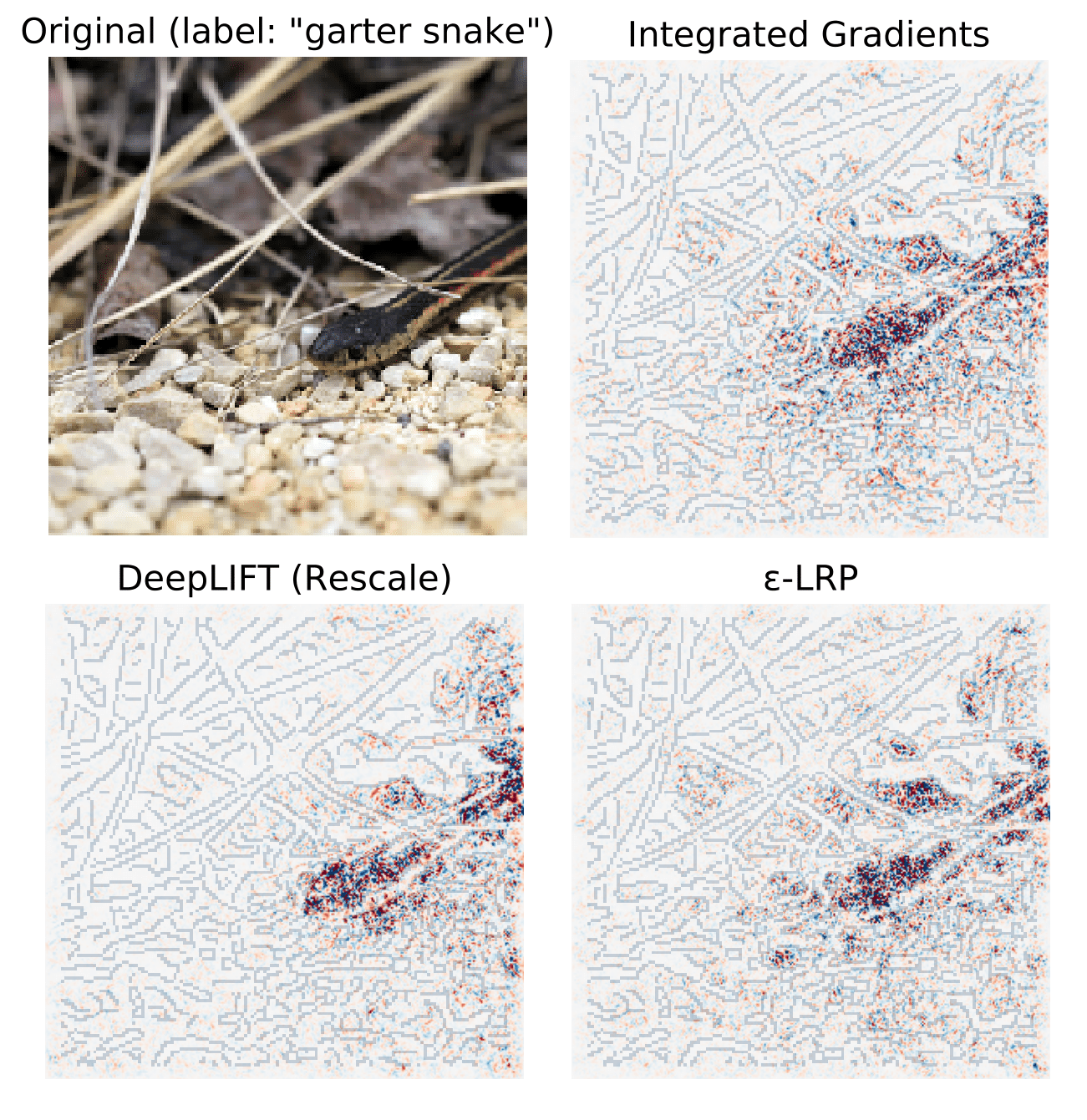}}
\par\end{centering}
\caption{\label{fig:Explainability-Algorithms-in}Explainability Algorithms
in the Image Recognition Domain (Taken from \cite{ancona2018towards})}
\end{figure}

The main contributions of this paper are as follows:
\begin{enumerate}
\item This is the first paper to demonstrate end-to-end adversarial examples
(that is, runnable malware) against malware classifiers that rely
on multiple types of interconnected and dependent features. This is
also the first such attack which not only adds features, but also
modify them while maintaining the malware functionality.
\item This is the first paper to discuss the usage of explainability by
adversaries, in order to both choose specific features to perturb
and maintain the minimal perturbation (that is, number of perturbed
features).
\item This is the first paper analyzing the concept of transferability of
explainability, using explainability of a white-box substitute model
to explain a black-box model, using a subset of the same indicative
features.
\end{enumerate}

\section{Related Work}

\subsection{\label{subsec:Explainability-Algorithms:}Explainable Machine Learning}

In this paper, we evaluate the usage of explainability algorithms
(which ranks the features by impact on the classification of a specific
sample, see Definition \ref{def:An-explainable-machine}) in order
to generate an adversarial example with a small perturbation size.
Several such algorithms have been introduced:

Integrated Gradients \cite{Sundararajan17}, computes the partial
derivatives of the output with respect to each input feature. Integrated
Gradients computes the average gradient while the input varies along
a linear path from a baseline $\bar{\boldsymbol{x}}$ to $\boldsymbol{x}$.
The baseline is defined by the user and often chosen to be zero. The
importance (also known as attribution) of $x_{i}$, the $i$-th element
in the vector $\boldsymbol{x}$:

\begin{equation}
R(x_{i})=(x_{i}-\bar{x_{i}})\intop_{\alpha=0}^{1}\frac{\partial C(\boldsymbol{\tilde{x}})}{\partial\tilde{x_{i}}}\mid_{\widetilde{\boldsymbol{x}}=\bar{\boldsymbol{x}}+\alpha(\boldsymbol{x}-\boldsymbol{\bar{x}})d\alpha}\label{eq:-4}
\end{equation}

, where $C(\boldsymbol{x})$ is the classifier prediction for $\boldsymbol{x}$.
Integrated Gradients satisfies an important property termed \emph{completeness}:
the attributions sum up to the target output minus the target output
evaluated at the baseline: $\sideset{}{_{i=1}^{N}}\sum R(x_{i})=C(\boldsymbol{x})-C(\boldsymbol{\bar{x}})$. 

Layer-wise Relevance Propagation (LRP) \cite{bach-plos15} is computed
with a backward pass on the network. Let us consider a quantity $r_{i}^{(l)}$
, called \textquotedbl relevance\textquotedbl{} of unit $i$ of layer
$l$. The algorithm starts at the output layer $L$ and assigns the
relevance of the target neuron c equal to the output of the neuron
itself and the relevance of all other neurons to zero (Equation \ref{eq:}). 

\begin{equation}
r_{i}^{(L)}=\begin{cases}
C_{i}(\boldsymbol{x}) & if\:unit\:i\:is\:the\:unit\:of\:interest\\
0 & otherwise
\end{cases}\label{eq:}
\end{equation}

The algorithm proceeds layer by layer, redistributing the prediction
score $C_{i}$ until the input layer is reached. One recursive rule
for the redistribution of a layer\textquoteright s relevance to the
following layer is the $\epsilon-rule$ described in Equation \ref{eq:-1},
where we define $z_{ji'}=w_{ji'}^{(l+1,l)}x_{i'}^{(l)}+b_{j}$ to
be the weighted activation of a neuron i' onto neuron $j$ in the
next layer and $b_{j}$ the additive bias of unit $j$. A small quantity
is added to the denominator of Equation \ref{eq:-1} to avoid numerical
instabilities. Once reached the input layer, the final importance
is defined as $R(x_{i})=r_{i}^{(1)}$ .

\begin{equation}
r_{i}^{(l)}=\sideset{}{_{j}}\sum\frac{w_{ji}^{(l+1,l)}x_{i}^{(l)}}{\sideset{}{_{i'}}\sum\left(z_{ji'}\right)+\epsilon*sign\left(\sideset{}{_{i'}}\sum\left(z_{ji'}\right)\right)}r_{j}^{(l+1)}\label{eq:-1}
\end{equation}

LRP together with the propagation rule described in Equation \ref{eq:-1}
is called $\epsilon-LRP$, analyzed in the remainder of this paper.

DeepLIFT \cite{Shrikumar17} proceeds in a backward fashion, similarly
to LRP. Each unit $i$ is assigned an attribution that represents
the relative effect of the unit activated at the original network
input $\boldsymbol{x}$ compared to the activation at some reference
input $\bar{\boldsymbol{x}}$ (Equation \ref{eq:-2}). 

\begin{equation}
r_{i}^{(L)}=\begin{cases}
C_{i}(\boldsymbol{x})-C_{i}(\bar{\boldsymbol{x}}) & if\:unit\:i\:is\:the\:unit\:of\:interest\\
0 & otherwise
\end{cases}\label{eq:-2}
\end{equation}

Reference values $\bar{z_{ji'}}$ for all hidden units are determined
running a forward pass through the network, using the baseline $\bar{\boldsymbol{x}}$
as input, and recording the activation of each unit. As in LRP, the
baseline is often chosen to be zero. The relevance propagation is
described in Equation \ref{eq:-1-1}. The attributions at the input
layer are defined as $R(x_{i})=r_{i}^{(1)}$ as for LRP.

\begin{equation}
r_{i}^{(l)}=\sideset{}{_{j}}\sum\frac{z_{ji}-\bar{z_{ji}}}{\sideset{}{_{i'}}\sum\left(z_{ji'}\right)-\left(\sideset{}{_{i'}}\sum\left(\bar{z_{ji'}}\right)\right)}r_{j}^{(l+1)}\label{eq:-1-1}
\end{equation}

In Equation \ref{eq:-1-1}, $\bar{z}_{ji'}=w_{ji'}^{(l+1,l)}\bar{x}_{i'}^{(l)}+b_{j}$
is the weighted activation of a neuron $i$ onto neuron $j$ when
the baseline $\bar{\boldsymbol{x}}$ is fed into the network. As with
Integrated Gradients, DeepLIFT was designed to satisfy \emph{Completeness}.
The rule described in Equation \ref{eq:-1-1} (\textquotedbl Rescale
rule\textquotedbl ) is used in the original formulation of the method
and it is the one we will analyze in the remainder of the paper.

\cite{ancona2018towards} formalized the above-mentioned methods in
similar terms and found some interesting similarities between them.
For instance, all methods are equivalent when the model behaves linearly,
e.g., when the network is very shallow.

SHAP (SHapley Additive exPlanation) \cite{NIPS2017_7062} values also
explain the output of a classifier $C$ as a sum of the effects $R(x_{i})$
of each feature being introduced into a conditional expectation. However,
unlike the above-mentioned methods, SHAP is a black-box method, which
doesn't require any knowledge about the architecture of the explained
network. Therefore, it cannot compute partial derivatives of the gradients
with respect to specific features. In order to evaluate the effect
missing features have on a model $C$ , it is necessary to define
a mapping between the missing features and the original function input
space. Therefore, In order to compute SHAP values, we define $C_{x}(S)=E\left(C(x)|x_{S}\right)$
where $S$ is the set of analyzed features, and $E\left(C(x)|x_{S}\right)$
is the expected value of the function conditioned on a subset S of
the input features. SHAP values combine these conditional expectations
with the classic Shapley values from game theory together to attribute
$R(x_{i})$ values to each feature:

\begin{equation}
R(x_{i})=\sideset{}{_{S\subseteq F-\{i\}}}\sum\frac{|S|!\left(|F-S|-1\right)!}{|F|!}\left[f_{x}\left(S\cup\{i\}\right)-f_{x}\left(S\right)\right]\label{eq:-1-1-1}
\end{equation}

, where $F$ is the set of all input features. Note that for non-linear
functions the order in which features are introduced matters. SHAP
values result from averaging over all possible orderings. Proofs from
game theory show this is the only possible consistent approach where
$\sideset{}{_{i=0}^{|F|}}\sum R(x_{i})=C(\boldsymbol{x})$. SHAP outperforms
other state-of-the-art black-box explainability algorithms, e.g.,
LIME \cite{DBLP:conf/ccs/GuoMXSWX18}.

Very few papers have been published about explainability in the cyber
domain. Gue et al. \cite{DBLP:conf/ccs/GuoMXSWX18} extends the LIME
algorithm to a variant called LEMNA, so it can be used for recurrent
neural networks (RNNs), commonly used in the cyber security domain
(but not in this paper and therefore it is not being evaluated here).
This is done using Gaussian mixture to better handle the non-linearities
of RNNs and fused lasso to merge features with similar importance
together.

Warnecke et al. \cite{DBLP:journals/corr/abs-1906-02108} have evaluated
the explainability of several of the algorithms mentioned above for
different datasets in the cyber security domain and concluded that
white-box explainability algorithms, such as integrated gradients
(Equation \ref{eq:-4}) and LRP (Equation \ref{eq:-1}) select features
which are both have a higher impact on the classification and provide
better human interpretable explanations, comparing to black-box explainability
algorithms such as SHAP (Equation \ref{eq:-1-1-1}) and LEMNA \cite{DBLP:conf/ccs/GuoMXSWX18}.

\subsection{\label{subsec:Adversarial-Examples}Adversarial Examples}

\cite{DBLP:journals/corr/SzegedyZSBEGF13} and \cite{Biggio2013}
formalize the search for adversarial examples as a similar minimization
problem:

\begin{equation}
\arg_{\boldsymbol{r}}\min C(\boldsymbol{x}+\boldsymbol{r})\neq C(\boldsymbol{x})\:s.t.\:\boldsymbol{x}+\boldsymbol{r}\in\boldsymbol{D}\label{eq:-3}
\end{equation}
The input \textbf{$\boldsymbol{x}$}, correctly classified by the
classifier $C$, is perturbed with \textbf{$\boldsymbol{r}$} such
that the resulting adversarial example \textbf{$\boldsymbol{x}+\boldsymbol{r}$}
remains in the input domain \textbf{$\boldsymbol{D}$}, but is assigned
a different label than \textbf{$\boldsymbol{x}$}. To solve Equation
\ref{eq:-3}, we need to transform the constraint $C(\boldsymbol{x}+\boldsymbol{r})\neq C(\boldsymbol{x})$
into an optimizable formulation. Then we can easily use the Lagrange
multiplier to solve it. To do this, we define a loss function $Loss()$
to quantify this constraint. This loss function can be the same as
the training loss, or it can be chosen differently, e.g., hinge loss
or cross entropy loss.

There are three types of adversarial examples generation methods:

Gradient based attacks - Those adversarial perturbation are generated
in the direction of the gradient, that is, in the direction with the
maximum effect on the classifier's output, e.g., FGSM \cite{Goodfellow14}
or JSMA \cite{Papernot2016}) Those attacks are effective but require
adversarial knowledge about the targeted classifier's gradients. Those
attacks can be conducted on the targeted model, if white-box knowledge
is available, or via the usage of a surrogate model, using the transferability
property (Appendix B) for a black-box attack.

Score based attacks - Those attacks are based on white-box knowledge
of the confidence score of the target classifier. The target classifier's
gradient can be numerically derived from confidence scores of adjacent
input points \cite{Chen2017} and then continue like gradient-based
attack, following the direction of maximum impact. A different approach
is using a genetic algorithm, where the fitness of the genetic variants
is defined in terms of the target classifier\textquoteright s confidence
score, to generate adversarial examples \cite{DBLP:conf/ndss/XuQE16}.

Decision based attacks - These attacks use only the label predicted
by the target classifier. \cite{brendel2018decisionbased} starts
from a randomly generated image classified as desired and then adds
perturbations that decrease the distance to the source class image,
while maintaining the target classification. The more noise you add,
the larger the chance of successfully modifying the classifier's decision.
However, the challenge is usually to use as little noise as possible,
since more noise might damage the original functionality of the input.
For instance, adding ``benign'' API calls to a malware might cause
a malware classifier to classify it as benignware, but might also
add performance overhead to the modified malware, due to the added
API calls. \cite{DBLP:conf/icml/IlyasEAL18} uses Natural Evolutionary
Strategies (NES) optimization to enable query-efficient gradient estimation,
which leads to generation of misclassified images like gradient based
attacks.

\subsubsection{\label{subsec:The-Transferability-Property}The Transferability Property}

Many black-box attacks (e.g., \cite{DBLP:conf/raid/RosenbergSRE18})
rely on the concept of \emph{adversarial example transferability},
presented in \cite{DBLP:journals/corr/SzegedyZSBEGF13}: Adversarial
examples crafted against one model are also likely to be effective
against other models. This transferability property holds even when
models are trained on different datasets. This means that the adversary
can train a \emph{substitute model}, which has similar decision boundaries
as the original model and perform a white-box attack on it. Adversarial
examples that successfully fool the surrogate model would most likely
fool the original model, as well (\cite{Papernot:2017:PBA:3052973.3053009}). 

The transferability between DNNs and other models such as decision
tree and SVM models was examined in \cite{DBLP:journals/corr/PapernotMG16}.
The reasons for the transferability are unknown yet, but a recent
study \cite{ilyas2019adversarial} suggests that adversarial vulnerability
is not ``necessarily tied to the standard training framework, but
is rather a property of the dataset (due to representation learning
of non robust features)'', which also clarifies why transferability
happens regardless of the classifier architecture.

\subsubsection{\label{subsec:End-To-End-Adversarial-Examples}End-To-End Adversarial
Examples against Malware Classifiers}

Attacks vary based on the amount of knowledge the adversary has about
the classifier being subverted: \emph{Black-Box attacks} requires
no knowledge about the model beyond the ability to query it as a black-box
(a.k.a. the \emph{oracle model}), i.e., inserting an input and getting
the output classification, while \emph{white-Box attacks} assume the
adversary has knowledge about the model architecture and even the
hyperparameters used to train the model. In this paper, we focus in
black-box attacks, which are a more realistic scenario in the cyber
security domain, in which security vendors don't reveal their architecture
to avoid being copied or bypassed.

Grosse et al. \cite{Grosse2017} presented a white-box attack against
an Android static analysis fully connected DNN malware classifier.
The static features used were from the AndroidManifest.xml file, including
permissions, suspicious API calls, activities, etc. The attack is
a discrete FGSM \cite{Goodfellow14} variant, which is performed iteratively
in two steps, until a benign classification is achieved: (1) Compute
the gradient of the white-box model with respect to the binary feature
vector $\boldsymbol{x}$. (2) Find the element in $\boldsymbol{x}$
whose modification from zero to one (i.e., only feature addition and
not removal) would cause the maximum change in the benign score, and
add this manifest feature to the adversarial example. Unlike our attack,
Grosse et al. only adds features and the APK format used is much simpler
than the PE format used in our attack, with no interdependent features.

Xu et al. \cite{DBLP:conf/ndss/XuQE16} generated adversarial examples
that bypass PDF malware classifiers, by modifying static PDF features.
This was done using an inference integrity genetic algorithm (GA),
where the fitness of the genetic variants is defined in terms of the
target classifier\textquoteright s confidence score. The GA is computationally
expensive and was evaluated against SVM, random forest, and CNN using
static PDF structural features. This attack requires knowledge of
both the classifier's features and the target classifier's confidence
score. Unlike our attack, Xu et al. only adds features and the PDF
format used is much simpler than the PE format used in our attack,
with no interdependent features.

Suciu et al. \cite{DBLP:conf/aaaifs/SuciuCJ18} implemented an attack
against MalConv, a 1D CNN, using the file's raw byte content as features
(Raff et al. \cite{DBLP:conf/aaai/RaffBSBCN18}). The additional bytes
are selected by the FGSM method and are inserted between the file's
sections. in a black-box manner by appending bytes from the beginning
of benign files. Unlike our attack, Suciu et al. use only a single
feature type (raw bytes) and only add features (and not modify features),
which is a less realistic scenario than ours.

Rosenberg et al. \cite{DBLP:conf/raid/RosenbergSRE18,DBLP:journals/corr/abs-1804-08778}
presented a black-box inference integrity attack that adds API calls
to an API call trace used as input to an RNN malware classifier in
order to bypass a classifier trained on the API call trace of the
malware. A GRU substitute model was created and attacked, and the
transferability property was used to attack the original classifier.
The authors extended their attack to hybrid classifiers combining
static and dynamic features, attacking each feature type in turn.
The target models were LSTM variants, GRUs, conventional RNNs, bidirectional
and deep variants, and non-RNN classifiers (including both feedforward
networks, like fully connected DNNs and 1D CNNs, and traditional machine
learning classifiers, such as SVM, random forest, logistic regression,
and gradient boosted decision tree). The authors presented an end-to-end
framework that creates a new malware executable without access to
the malware source code. Unlike our attack, Rosenberg et al. only
adds features, not modify them and only uses two feature types, which
are independent, making this case less challenging and realistic than
ours.

In Anderson et al. \cite{DBLP:journals/corr/abs-1801-08917}, the
features used by the gradient boosted decision tree classifier included
PE header metadata, section metadata, and import/export table metadata.
A black-box attack which trains a reinforcement learning agent was
presented. The agent is equipped with a set of operations (such as
packing) that it may perform on the PE file. The reward function was
the evasion rate. Through a series of games played against the target
classifier, the agent learns which sequences of operations are likely
to result in detection evasion for any given malware sample. The differences
from our attack are: (1) Our attack also achieve higher attack effectiveness
then Anderson et al. with less adversary's knowledge (different classifier's
type and architecture, training set and feature subset), due to our
usage of transferability of explainability. (2) The attack in Anderson
et al. uses whole-PE transformations like packing, which increases
the chances of the generated malware to be detected by anomaly detection
methods (e.g., \cite{DBLP:journals/scn/ZakeriDA15}). (3) Our attack
effectiveness is 37\%, as opposed to less than 25\% for Anderson et
al. on the same dataset and using the same attacked classifier. Note
that both attacks have a lower effectiveness than image-based attacks
due to the challenges mentioned in Section \ref{subsec:The-Challenges-in}. 

This paper is the first to present end-to-end PE structural features
adversarial examples, which, unlike previous attacks, include feature
modification (and not just addition) without harming the malware functionality
and interdependent features.

\subsubsection{Using Explainable ML in Adversarial Scenarios}

Several papers have tried to leverage the usage of explainability
to detect adversarial examples, although none of them are in the cyber
security domain. 

Tau et al. \cite{DBLP:conf/nips/TaoMLZ18} generated a mapping to
the neurons critical for specific attributes and amplified the activation
of those neurons to make the classifier more robust to adversarial
attacks. Carlini \cite{DBLP:journals/corr/abs-1902-02322} demonstrated
that this robust classifier is still vulnerable to the Carlini and
Wagner attack.

Fidel et al. \cite{DBLP:journals/corr/abs-1909-03418} used the SHAP
values computed for the internal layers of a DNN classifier to discriminate
between normal and adversarial inputs. Amosy et al. \cite{chechik2020using}
used a similar approach. 

To the best of our knowledge, our paper is the first to leverage explainability
algorithms from the adversary side, to generate and facilitate adversarial
attacks (as opposed to detecting them).

\section{Methodology}

\subsection{\label{subsec:Threat-Model}Threat Model}

The adversary's goal is to modify a malware executable for it to bypass
a multi-feature types malware classifier without harming the executable's
functionality (Section \ref{subsec:Executable-(Malicious)-Functiona}),
that is, generating an end-to-end malware adversarial example. In
this paper we limit ourselves to static features, that is, features
that can be extracted from the file without running it. Static features
are the malware file's content and properties (e.g., the file's size)
either as raw bytes or pre-processed to parse them in the way used
by the operating system loads them, termed PE structural features.
Raw-byte features require a very long training process and current
state-of-the-art GPU hardware usually limits the file size which can
be classified using such classifier (e.g. \cite{DBLP:conf/aaai/RaffBSBCN18}),
making this a non realistic use case. Using dynamic features (e.g.,
executed API calls in Rosenberg et al. \cite{DBLP:conf/raid/RosenbergSRE18})
is also less common use case, since it requires a sandbox environment
in order to avoid running a malware on the computer we want to protect,
which might harm it. We therefore decided to focus on PE structural
features, which are used by real-world classifiers. We assume the
adversary has no knowledge or access to the attacked malware classifier,
e.g., the classifier type, architecture or training set (a black-box
attack, as defined in Section \ref{subsec:End-To-End-Adversarial-Examples}).
Our attack would be a decision-based attack (by the definition in
Section \ref{subsec:Adversarial-Examples}), as this is the most realistic
scenario. We do assume the adversary can figure out \textbf{some}
of the features used by the attacked malware classifier, but not all
of them. This is a common case in cyber security, especially with
static features, where many classifiers are using similar PE structural
features, (e.g., \cite{Saxe2015,DBLP:journals/corr/abs-1801-08917,DBLP:journals/corr/abs-1804-04637})
but the exact subset of features is unknown. 

\subsection{\label{subsec:Generating-an-End-to-End}Generating an End-to-End
Adversarial Example}

In order to evade detection by the malware classifier, the adversary
is using the method specified in Algorithm \ref{alg:End-to-End-PE-Structural}.

\begin{algorithm}[t]
\caption{\label{alg:End-to-End-PE-Structural}End-to-End PE Structural Features-Based
Adversarial Example Generation}

\begin{enumerate}
\item Train a substitute neural network model on a training set and features
believed to accurately represent the attacked malware classifier.
\item Select a malware binary he/she wants to bypass the attacked malware
classifier.
\item Use explainable machine learning algorithm (see Definition \ref{def:An-explainable-machine})
to get a list of features importance for the classification of the
malware on the substitute model (see Section \ref{subsec:Transferability-of-Explainabilit}).
\item From those features, choose those easier to modify (see Section \ref{subsec:Easily-Modifiable-PE}).
\item \label{enu:Modify-each-easily}Modify each ``easily modifiable''
feature using the list of predefined feature values (see Section \ref{subsec:Easily-Modifiable-PE}),
selecting the value that result in the lowest confidence score. Repeat
until a benign classification is achieved by the target black-box
malware classifier.
\end{enumerate}
\end{algorithm}

In this method, we use the following definition:
\begin{defn}
\label{def:An-explainable-machine}An explainable machine learning
algorithm $A(m,\boldsymbol{v})$ takes as arguments a machine learning
model $m$ and a sample's vector $\boldsymbol{v}$ and returns a vector
of $length(\boldsymbol{v})$ values which represent the weights of
impact of the features in $\boldsymbol{v}$, such that a higher weight
indicates a more impactful feature for classifying the vector $\boldsymbol{v}$
by the model $m$. 
\end{defn}
This method is relying on two assumptions, evaluated in the following
subsections:

(1) The most important features in the attacked malware classifier
would be similar to those of the substitute model, and they would
also be found by an explainability algorithm. Thus, modifying these
features in the method mentioned above would affect the attacked malware
classifier as well. Detailed in Section \ref{subsec:Transferability-of-Explainabilit-1},
and (2) The adversary can modify the malware binary without harming
its functionality. Detailed in Section \ref{subsec:End-to-End-Feature-Modification}.

\subsection{\label{subsec:Transferability-of-Explainabilit-1}Transferability
of Explainability}

The concept of transferability of explainability is defined as follows:
\begin{defn}
\label{def:Given-two-different}Given two different models, $m_{1}$
and $m_{2}$ with different classifier type and architecture trained
on a similar dataset and input features list, the output of an explainable
machine learning algorithm (see Definition \ref{def:An-explainable-machine})
would be similar for $m_{1}$ and $m_{2}$.
\end{defn}
Note that this definition is different from adversarial examples transferability:
Adversarial examples transferability (see Section \ref{subsec:The-Transferability-Property})
is the concept of an adversarial example generated to fool one classifier
is also effective against another classifier. Transferability of explainability
means that the feature group indicated to have a high impact on a
specific sample classification on one model would be similar to the
list of the same explainability algorithm on another model. We argue
that this holds true regardless of the classifier type, architecture,
training set or even explainable algorithm. The only requirement is
that the features used by both classifiers need to be similar enough
(otherwise impactful features in one model are meaningless in the
other model) - but not identical. A visual example of the transferability
of explainability can be seen in Figure \ref{fig:Explainability-Algorithms-in}.
We see that three different explainable algorithms (Integrated Gradients
\cite{Sundararajan17}, DeepLIFT \cite{Shrikumar17} and Layer-wise
Relevance Propagation (LRP) \cite{bach-plos15}) highlight similar
features as the most important to classify a gartner snake image (mainly,
pixels in the snake's head area).

This concept is especially important for multi-feature types malware
classifiers: On the one hand, the adversary is unaware of the attacked
classifier architecture, so using transferability is essential. On
the other hand, modifying too many features might cause the adversarial
example to be caught by anomaly detectors (e.g., \cite{DBLP:journals/scn/ZakeriDA15}).
Therefore, a small perturbation (that is, modifying a small amount
of features) is desired.

Using transferability of explainability to generate adversarial examples
is also usable in scenarios where using transferability of adversarial
examples (as done in the adversarial attacks mentioned in Section
\ref{subsec:Adversarial-Examples} for a black-box scenario) is not:

(1) When there are dependent features, which modification requires
the modification of other features for the file to continue being
runnable, as in the case of PE structural features, discussed in this
paper (see Section \ref{subsec:Feature-Modification-Challenges}).
In this case, it's very hard to take into account which features need
to be modified to keep a small perturbation automatically.

(2) When some features are harder to modify than others, which is
very hard to take into account in any mathematical form \cite{KATZIR2018419},
which can be used automatically.

In contrast, a manual modification of the features by their order
of impact is a preferable approach to keep the perturbation small.
Such cases might be the reason why there are no adversarial attacks
published on end-to-end PE structural features yet (see Section \ref{subsec:End-To-End-Adversarial-Examples}).

\subsection{\label{subsec:End-to-End-Feature-Modification}End-to-End Feature
Modification for PE Structural Features}

In this paper we are focusing on malware running on Windows OS, since
most malware target the Windows OS. Thus, we focus on the static features
of the executable format on Windows, named portable executable (PE).

\subsubsection{\label{subsec:PE-Structural-Features-1}PE Structural Features Overview}

In this section we discuss the features that exist in the dataset
used in this paper, Ember. However, as mentioned in Section \ref{subsec:Threat-Model},
many classifiers are using similar PE structural features, so this
description is valid to all of them. A more detailed list of the Ember
model features appear in \cite{DBLP:journals/corr/abs-1804-04637}.
Here we will only describe some major key points regarding the feature
types used in the model. Some features are naive values extracted
from the PE header with no modification at all and some features are
engineered, for example, string features which count occurrences of
Windows path strings. On a high-level, the PE file itself is composed
of the PE header, sections and overlay. The PE header in turn is composed
of various fields and additional headers, e.g., DOS header and Optional-Header.
The sections are either code sections (machine instructions), data
sections (holding variables) and resource sections (holding embedded
fonts, images, etc). The overlay is defined as any addition to the
PE file that is not defined in the various fields in the headers and
therefore not loaded into the process memory. The PE structure has
a lot of flexibility. For example, there are numerous entries to describe
a section but only few are necessary. Moreover, some values differ
when loaded into memory and when viewed statically as they appear
in the file. For example, various offsets and relocation are resolved
by the Windows process loader and the values are modified during process
mapping preparation before executing it.

\paragraph{Feature types description}

The different feature types in Ember model are:

(1) Byte histogram - A total of 256 features that describe the byte
value histogram in the entire file.

(2) Byte entropy histogram - A total of 256 features that roughly
approximates the joint distribution of byte value and local entropy
(see \cite{Saxe2015}).

(3) String related features - A total of 104 string related features,
s.a., total count, average string length, Windows path count, etc.
96 features of this group are printable character distribution. We
will mostly focus on this subset of the string features.

(4) General information features - A total of 10 features describing
properties of the PE file, e.g., import and export functions count,
has relocation table, has resource table, etc.

(5) COFF header features - A total of 62 features describing values
from the COFF header. Out of these, 50 features use the hash trick
with 10 buckets over 5 fields in the COFF header.

(6) Section features - A total of 255 features describing values from
the section headers. Out of these, 5 are counters and the other 250
features are hash trick with 50 buckets over 5 fields.

(7) Imports features - A total of 1280 features. All use the hash
trick, 256 hash buckets for import library names and 1024 for imported
function names.

(8) Exports features - A total of 128 features. All use the hash trick
for exported function names.

(9) Data directories features - A total of 30 features, describing
values of size and virtual size for 15 data directories entries present
in the PE.

\subsubsection{\label{subsec:Feature-Modification-Challenges}Feature Modification
Challenges}

Our goal is to change the model prediction for a given PE file, while
not harming the functionality of the PE (Section \ref{subsec:Executable-(Malicious)-Functiona}).
It is easy to see that some feature types are interdependent, for
example, modifying some of the string features will affect the byte
histogram and byte entropy histogram. Other features may prove to
be difficult to near impossible to modify. For example, the hash buckets
values can be affected by inserting the relevant value and making
sure that the generated hash falls in the required bucket, however
it is not always possible to use the resulting value(s) in every field.
With strings it may be simple but changing certain values in the header
might render the PE file as non executable. It is important to notice
that we can only affect feature values in a limited way. Some values
cannot be altered, for example if an import function is being used,
we will not able to alter the value of the hash bucket to other than
a non-zero value.

\subsubsection{\label{subsec:Easily-Modifiable-PE}Easily Modifiable PE Structural
Features}

Analyzing the feature types mentioned above, we notice that there
are types of features which can be modified easily without affecting
the execution of the PE file as a process when loaded into memory
by the OS (and thus the modified PE file's functionality). 

One such example is the printable character distribution (Section
\ref{subsec:PE-Structural-Features-1}). Considering the character
distribution features, we needed to not only keep the values that
resulted in the greater shift in prediction score but also recalculate
the entire distribution of characters in the file and generate a buffer
that when appended to the original file will tilt the distribution
accordingly. That buffer can be appended to the end of the file (overlay)
but we also chose to insert it as a new section to the PE file, therefore
making it a little less trivial to explain the prediction score difference
change by an examining eye. The buffers themselves were bound by size
but in most test cases we only need several hundred kilo-bytes of
data to successfully shift the prediction.

Other features we can modify come mostly from the PE header and its
composing headers and values. Listed below is a short description
of such features: (1) PE COFF Header timedate stamp - 4 bytes that
hold the linkage time of the executable. It has no affect whatsoever
on the execution of the PE file. (2) PE CLR Runtime Size - A field
that describes the .Net runtime size, used only by the .NET VM when
the pe is linked with mscoree.dll (3) PE CLR Runtime Virtual Address
- A field that describes the .Net header virtual address, used only
by the .NET VM when the pe is linked with mscoree.dll . The features
we perturbed can be seen in our git repository.

For each feature, our attack (Algorithm \ref{alg:End-to-End-PE-Structural},
step \ref{enu:Modify-each-easily}) iterates over a list of predetermined
features and alters each of them according to a predefined set of
values that matches the feature type possible values. The predefined
list of values per feature serves two purposes:

(1) It limits the amount of iterations it takes to complete the brute-force
for a specific file, and (2) It verifies that the value ranges fits
the feature. For instance, the header feature MajorOperatingSystemVersion
is the minimum version of the operating system required to use this
executable. Putting a large value (e.g., Windows 10) might prevent
the Windows loader from running the modified file on relevant machines
(e.g., Windows 7 hosts).

\section{Experimental Evaluation}

\subsection{\label{subsec:Dataset-and-Classifiers}Dataset and Classifiers}

As mentioned in Section \ref{subsec:PE-Structural-Features-1}, we
used the Ember dataset. It is thoroughly described in \cite{DBLP:journals/corr/abs-1804-04637},
and is the state-of-the-art dataset of 1M malware and benign-ware,
equally distributed. We split the dataset into a training-set of 300K
malware and 300K benignware and a test set of 200K malware and 200K
benignware.

As the target classifier, we used the gradient boosted decision tree
(GBDT) classifier used in \cite{DBLP:journals/corr/abs-1804-04637},
which outperformed state-of-the-art raw features model \cite{DBLP:conf/aaai/RaffBSBCN18}.
This classifier input is a vector of 2381 Ember's PE structural features
and its output is a binary classification: malicious or benign file.
It is trained using \href{https://github.com/microsoft/LightGBM}{LightGBM}
with 100 trees and 31 leaves per tree.

As a substitute model, trained by the adversary, we used an architecture
similar to the one used in Saxe et al. \cite{Saxe2015}, which also
uses PE structural features. It contains two hidden dense layers with
128 neurons, ReLU activation functions and dropout rate of 0.2, followed
by a final dense layer with a sigmoid activation layer. The input
and output are identical to the attacked malware classifier. The substitute
model was trained with \href{https://github.com/keras-team/keras}{keras},
using a \href{https://github.com/tensorflow/tensorflow}{tensorflow}
backend.

\subsection{\label{subsec:Transferability-of-Explainabilit}Transferability of
Explainability for PE Structural Features Based Multi-Feature Type
Malware Classifiers}

We want to evaluate the concept of transferability of explainability
(Section \ref{subsec:Transferability-of-Explainabilit-1}) in our
setting. Mainly, we want to show that the most impactful features
in the substitute model are similar to those in the attacked classifier,
allowing the adversary to have only black-box access to the attacked
malware classifier. We therefore want a measure of the correspondence
between two rankings. We evaluated three metrics: We used Kendall's
tau \cite{Kendall45} to compare between the feature rankings of different
classifiers. Kendall's tau between two rankins, $\boldsymbol{r_{1}}$
and $\boldsymbol{r_{2}}$, with the same number of elements, is defined
as:

\begin{equation}
\tau(\boldsymbol{r_{1}},\boldsymbol{r_{2}})=\frac{P-Q}{\sqrt{(P+Q+T)*(P+Q+U)}}
\end{equation}

, where $P$ is the number of agreeable (concordant) pairs, $Q$ the
number of non-agreeable (discordant) pairs, $T$ the number of ties
(two elements are the same) in $\boldsymbol{r_{1}}$, and $U$ the
number of ties in $\boldsymbol{r_{2}}$. If a tie occurs for the same
pair in both $\boldsymbol{r_{1}}$ and $\boldsymbol{r_{2}}$, it is
not added to either $T$ or $U$. Values close to 1 indicate strong
agreement (and transferability), values close to -1 indicate strong
disagreement (and lack of transferability) between the ranking orders
of the two classifiers. Kendall's tau has advantages over other rank
metrics such-as Spearman\textquoteright s rho: The distribution of
Kendall\textquoteright s tau has better statistical properties, and
the interpretation of Kendall\textquoteright s tau in terms of the
probabilities of observing the agreeable and non-agreeable pairs is
very direct. 

We also evaluated a variant of the weighted Kendall's tau \cite{Vigna15}
metric. The weighted Kendall's tau is a weighted version of Kendall\textquoteright s
tau in which exchanges of high weight are more influential than exchanges
of low weight. We used the metric: $\tau w_{pos}(\boldsymbol{r_{1}},\boldsymbol{r_{2}})$
use additive hyperbolic weighing, that is, rank $r$ is mapped to
weight $1/(r+1)$, which has been shown to provide the best balance
between important and unimportant elements \cite{Vigna15}. Note that
$\tau w_{pos}(\boldsymbol{r_{1}},\boldsymbol{r_{2}})$ gives more
weight to important features, while $\tau(\boldsymbol{r_{1}},\boldsymbol{r_{2}})$
weighs all features the same.

We evaluated four different state-of-the-art explainability algorithms.
We focused on white-box explainability algorithms on the substitute
model because \cite{DBLP:journals/corr/abs-1906-02108} showed that
this yields better results for cyber security classifiers. Therefore,
we evaluated: Integrated Gradients \cite{Sundararajan17}, DeepLIFT
\cite{Shrikumar17} and $\epsilon-LRP$ \cite{bach-plos15} (see Section
\ref{subsec:Explainability-Algorithms:}). We compared these methods
to the black-box explainability algorithm SHAP \cite{NIPS2017_7062}.
Each of these algorithms' feature ranks were used to compare between
our substitute model and the attacked malware classifier (detailed
in Section \ref{subsec:Dataset-and-Classifiers}), on various levels
of knowledge of the adversary:

(1) The substitute model has the same training set and features as
the attacked classifier but a different architecture.

(2) The substitute model has the same features as the attacked classifier
but a different training set and architecture.

(3) The substitute model has different feature subset, training set
and architecture.

In our evaluation, the different architectures of the substitute and
attacked malware classifiers are specified in Section \ref{subsec:Dataset-and-Classifiers}.
Different training sets were obtained by randomly dividing the Ember
training set (specified in Section \ref{subsec:Dataset-and-Classifiers})
into two equal-sized training sets of 300K samples (150K malicious,
150K benign) each, used by the attacked malware classifier and the
other by the substitute model. This means there wasn't even a single
shared sample, but the samples were from the same distribution (e.g.,
same prominent malware families were represented in both training
sets), as expected in real-world cases. The accuracy of the attacked
classifier on the test set is 97.57\% (95.55\% for the substitute
model). When training the same classifier on half the training set,
the accuracy reduces to 97.23\% (90.81\% for the substitute model),
so the drop in accuracy is not big. Different features were obtained
by randomly picking two subsets of 1190 of Ember's 2381 features (50\%
of the features). This resulted in 32\% of the 2381 features being
used by both attacked and substitute classifiers, resulting in a reduced
attack surface, because only those features can be modified and affect
the attacked model. This emulates the real world scenario where many
malware classifiers use similar feature subsets (see Section \ref{subsec:Threat-Model}).
When training the attacked classifier on half the training set using
50\% of the features (the same subset used in our attacks), the accuracy
reduces to 96.77\% (88.94\% for the substitute model), so the drop
in accuracy is not big here either. The false positive rate of all
three models was about 1\%. The substitute model achieved similar
false positive rate on the test set in all evaluated scenarios mentioned
above. The results are shown in Table \ref{tab:Transferability-of-Explainabilit}.
When a random subset was made (whether in the training set or the
features), the same random permutation was used in all use cases in
the table. We used the explainability algorithms implementation in
\href{https://github.com/marcoancona/DeepExplain}{DeepExplain}. We
used the \href{https://docs.scipy.org/doc/scipy/reference/generated/scipy.stats.kendalltau.html}{Kendall's tau}
and \href{https://docs.scipy.org/doc/scipy/reference/generated/scipy.stats.weightedtau.html\#scipy.stats.weightedtau}{weighted Kendall's tau}
implementation of \href{https://github.com/scipy/scipy}{scipy}. The
baseline $\bar{x}$ for the integrated gradients and DeepLIFT methods
was chosen as a vector of zeros.

\begin{table}[t]
\caption{\label{tab:Transferability-of-Explainabilit}Transferability of Explainability
by ($\tau w_{pos}(\boldsymbol{r_{1}},\boldsymbol{r_{2}})$ | $\tau(\boldsymbol{r_{1}},\boldsymbol{r_{2}})$)
Metrics}

\centering{}%
\begin{tabular}{|>{\centering}p{0.12\paperwidth}|>{\centering}p{0.07\paperwidth}|>{\centering}p{0.07\paperwidth}|>{\centering}p{0.07\paperwidth}|}
\hline 
Explainability Algorithm\textbackslash Adversary's Knowledge & Same training set and features & Different training set, same features & Different training set and feature subset\tabularnewline
\hline 
\hline 
Integrated Gradients \cite{Sundararajan17} & 0.998 | 0.991 & 0.969 | 0.832 & 0.928 | 0.660\tabularnewline
\hline 
\hline 
DeepLIFT \cite{Shrikumar17} & 0.998 | 0.992 & 0.962 | 0.797 & 0.923 | 0.641\tabularnewline
\hline 
\hline 
$\epsilon-LRP$\cite{bach-plos15} & 0.998 | 0.992 & 0.963 | 0.801 & 0.926 | 0.652\tabularnewline
\hline 
\hline 
SHAP \cite{NIPS2017_7062} & 0.997 | 0.989 & 0.981 | 0.889 & 0.934 | 0.682\tabularnewline
\hline 
\hline 
Random Feature Ranking & 0.114 | 0.001 & 0.123 | 0.013 & 0.018 | 0.007\tabularnewline
\hline 
\end{tabular}
\end{table}

We see that all the explainability algorithms operating on the substitute
model reach a much better correlation (or ranking similarity) between
the feature importance in the substitute and attacked models than
random feature ranking. Unlike \cite{DBLP:journals/corr/abs-1906-02108}
, we see that SHAP is on par with the other, white-box explainability
algorithms, possibly due to the different metric we used, which fit
our use case of selecting the features to perturb. This high correlation,
very close to 1.0, shows that the transferability of explainability
does exist in our setting, and the most important features explaining
the substitute model's classification can just as well explain (and
affect) the classification of the attacked model. We see that, as
expected, the correlation (or ranking similarity) between the feature
rankings decreases, as expected when the attacker has less information
about the training set or the features. However, even under those
constraints, the correlation is relatively high. We also see that
$\tau w_{pos}(\boldsymbol{r_{1}},\boldsymbol{r_{2}})$ provides higher
correlation than $\tau(\boldsymbol{r_{1}},\boldsymbol{r_{2}})$ in
all cases. This shows that while the ranking might be different, the
most important features (which are what important for our attack,
described in Algorithm \ref{alg:End-to-End-PE-Structural}) are still
ranked similarly.

\subsection{\label{subsec:PE-Structural-Features}PE Structural Features Based
Multi-Feature Type Malware Classifier End-to-End Adversarial Examples}

In order to measure the performance of an attack, we consider two
factors:

The \emph{attack effectiveness} is the percentage of malicious samples
which were correctly classified by the attacked malware classifier,
for which the end-to-end adversarial example generated by Algorithm
\ref{alg:End-to-End-PE-Structural} was misclassified as benign by
the attacked malware classifier.

The average \emph{perturbation size} is the average number of features
(out of the Ember dataset's total of 2381) that were perturbed before
the attack was successful. The adversary aims to minimize it in-order
to evade detection by, e.g., anomaly detection classifiers that recognize
anomalous PE structure \cite{DBLP:journals/scn/ZakeriDA15}.

We generated end-to-end adversarial examples to all the malicious
samples in the Ember test set which were correctly detected by the
attacked malware classifier (a total of: 97570 samples). The results
of the attack are shown in Table \ref{tab:Adversarial-Examples-Success}.

\begin{table}[b]
\caption{\label{tab:Adversarial-Examples-Success}Adversarial Examples Success
Rate and Average Number of Modified Features}

\centering{}%
\begin{tabular}{|>{\centering}p{0.14\paperwidth}|>{\centering}p{0.08\paperwidth}|>{\centering}p{0.08\paperwidth}|>{\centering}p{0.08\paperwidth}|}
\hline 
Explainability Algorithm\textbackslash Adversary's Knowledge & Same training set and features & Different training set, same features & Different training set and feature subset\tabularnewline
\hline 
\hline 
Integrated Gradients \cite{Sundararajan17} & 37.71\% | 3.46 & 34.63\% | 3.83 & 34.19\% | 3.88\tabularnewline
\hline 
\hline 
DeepLIFT \cite{Shrikumar17} & 37.71\% | 3.45 & 34.38\% | 3.86 & 34.01\% | 3.91\tabularnewline
\hline 
\hline 
$\epsilon-LRP$\cite{bach-plos15} & 37.70\% | 3.46 & 34.42\% | 3.86 & 34.13\% | 3.90\tabularnewline
\hline 
\hline 
SHAP \cite{NIPS2017_7062}  & 37.66\% | 3.47 & 35.06\% | 3.79 & 34.42\% | 3.87\tabularnewline
\hline 
\hline 
Random Feature Ranking & 36.36\% | 4.10 & 11.54\% | 3.63 & 0.11\% | 7.91\tabularnewline
\hline 
\end{tabular}
\end{table}

We see that the regardless of which explainability algorithm we use,
the attack effectiveness is very similar. While reducing the attacker
knowledge brings to lower attack accuracy and higher number of modified
features, the attack is still effective for more than third of the
malware, even with different training sets and when only 32\% of the
features to perturb are known. Selecting the most important features
of a random feature ranking (which is equal to random feature selection),
brings interesting results. When the attacker has full knowledge about
the attacked classifier, the attack effectiveness is not much worse
than using the substitute model for selecting the important features
to perturb. However, when the attacker has less knowledge (as usually
happens in real-world scenarios), the random selction attack effectiveness
is significantly reduced. This shows the power and importance of using
the explainability approach in real-world scenarios.

\section{Conclusions and Future Work}

In this paper, we present a method to generate end-to-end multi-feature
types adversarial examples for PE malware classifiers, using explainability
algorithms to decide which features to modify. Our method is the first
to tackle the challenging task of generating end-to-end adversarial
examples of PE structural features, allowing not only feature addition
but also feature modification.

Our evaluation demonstrates that explainability is a dual edged sword,
which can also be leveraged by adversaries. When considering the call
to generate more explainable models, which decisions can be interpreted
by humans \cite{DBLP:journals/corr/abs-1811-10154} , one should take
into account its negative effects, such as making adversarial examples
less challenging in certain situations, as presented in this paper.

Our future work will include improving the query-efficiency of our
attack (in sense of queries to the attacked malware classifier), in
order to make it useful to attack cloud-based classifiers, by using
gradient-based approaches (e.g., JSMA \cite{Papernot2016}) over the
substitute model in order to find the optimal feature modification
out of the initial predetermined list. We would also research the
detection and defense methods against such attacks, for instance,
anomaly detection classifiers that recognize anomalous PE structure.

\bibliographystyle{IEEEtran}
\bibliography{IJCNN-compact}

\end{document}